\documentclass[12pt]{iopart}

\usepackage{iopams} 
\usepackage{amsmath}
\usepackage{xcolor}
\usepackage{enumerate}
\usepackage{comment}
\usepackage{appendix}

\begin{document}

\title{A Relativistic Scalar Model for Fractional Interaction between Dark Matter and Gravity}

\author{Francesco Benetti$^{1,2}$, Andrea Lapi$^{1,2,3,4}$, Giovanni Gandolfi$^{5,2}$, Stefano Liberati$^{1,2,3}$}
\address{
\begin{enumerate}[1 ]
    \item Scuola Internazionale Superiore Studi Avanzati (SISSA), Physics Area, Via Bonomea 265, 34136 Trieste, Italy
    \item Institute for Fundamental Physics of the Universe (IFPU), Via Beirut 2, 34014 Trieste, Italy
    \item Istituto Nazionale Fisica Nucleare (INFN), Sezione di Trieste, Via Valerio 2, 34127 Trieste, Italy
    \item Istituto di Radio-Astronomia (IRA-INAF), Via Gobetti 101, 40129 Bologna, Italy
    \item Dipartimento di Fisica e Astronomia, Univ. di Padova, Vicolo dell’Osservatorio, 3, 35122 Padova, Italy 
\end{enumerate}}
\ead{fbenetti@sissa.it, lapi@sissa.it}
\vspace{10pt}
\begin{indented}
\item[]
\end{indented}

\begin{abstract}
    In a series of recent papers we put forward a ``fractional gravity'' framework striking an intermediate course between a modified gravity theory and an exotic dark matter (DM) scenario, which envisages the DM component in virialized halos to feel a non-local interaction mediated by gravity. The remarkable success of this model in reproducing several aspects of DM phenomenology motivates us to look for a general relativistic extension. Specifically, we propose a theory, dubbed Relativistic Scalar Fractional Gravity or RSFG, in which the trace of the DM stress-energy tensor couples to the scalar curvature via a non-local operator constructed with a fractional power of the d'Alembertian. We derive the field equations starting from an action principle, and then we investigate their weak field limit, demonstrating that in the Newtonian approximation the fractional gravity setup of our previous works is recovered. We compute the first-order post-Newtonian parameter $\gamma$ and its relation with weak lensing, showing that although in RSFG the former deviates from its GR values of unity, the latter is unaffected. We also perform a standard scalar-vector-tensor-decomposition of RSFG in the weak field limit, to highlight that gravitational waves propagate at the speed of light, though also an additional scalar mode becomes dynamical. Finally, we derive the modified conservation laws of the DM stress energy tensor in RSFG, showing that a new non-local force emerges, and hence that the DM fluid deviates from the geodesic solutions of the field equations.
\end{abstract}

\vspace{2pc}
\noindent{\it Keywords}: dark matter, non-local gravity
\maketitle

\section{Introduction}\label{sec|intro}

In recent years there has been an increased interest in non-local gravity theories, envisaging that gravity may feature a self-interaction mediated by non-local operators. Among these theories the ones involving an infinite number of derivatives are arguably the most promising, since some of them have been shown to yield singularity-free, ghost-free (and thus unitary) and renormalizable theories of quantum gravity \cite{beneito_Calcagni,calcagni2023renormalizability,Maggiore1}.

The actions of these theories are generically constructed adding to the Einstein-Hilbert one of General Relativity (GR) all the possible terms involving the scalar curvature, the Ricci and Riemann or Weyl tensors, with non-local infinite derivatives operators sandwiched between them. In order to preserve general covariance, these operators are usually taken to be functions of the d'Alembertian $\Box = \nabla_{\mu}\nabla^{\mu}$ or of its inverse $\Box^{-1}$. Since the d'Alembertian has dimensions of length$^{-2}$, the former class of theories offer a completion of GR in the ultraviolet regime (i.e., small scales), being able to smooth the black holes and Big Bang singularities that are inevitable predictions of classical GR \cite{Mazumdar1,Biswas_2010,buoninfante_2022black,Penrose65}, while the latter modify gravity in the infrared regime (i.e., large scales) and can lead to interesting cosmological consequences. In fact, it has been shown that models of this kind can give rise to the late time cosmic accelerated expansion of the Universe without invoking any dark energy component \cite{Foffa_2014,Capozziello_2021,Maggiore_2014}.

In a nutshell, that infinite derivatives theories are non-local can be understood with the following basic argument: in order to calculate the first derivative of a standard function, its value at two infinitesimally close points must be known; similarly, the $n$-th derivative requires knowledge of the function value at $n+1$ different points. This means that infinitely many derivatives of a field require knowledge of its value throughout all spacetime. Another way to see this is by looking at the action of the translation operator of a function $e^{\epsilon\frac{\rm d}{{\rm d}x}}f(x) = f(x+\epsilon)$. Since $e^{\epsilon\frac{\rm d}{{\rm d}x}} = \sum_{n}\frac{\epsilon^n}{n!}\left(\frac{\rm d^n}{{\rm d}x^n}\right)$, this suggests that actions involving infinitely many derivatives can be recast in terms of fields translated at different points in spacetime. 

On the other hand, inverse powers of the d'Alembertian lead to manifestly non-local theories, since $\Box^{-1}$ can be represented as an integral kernel acting via the Green function of the $\Box$ operator:
\begin{equation}
(\Box^{-1}f)(x) = \int{\rm d}^4y\; \sqrt{-g(y)}\, G(x,y)\, f(y)\;.
\end{equation}
To different choices of Green functions correspond alternative definitions of these inverse operators. In particular, requiring causality, meaning that an event at a spacetime point could only be influenced by events in its past but not in its future, amounts to select the retarded green function $G_{\rm ret}(x-y)$. Note that this choice can be made only a posteriori, by replacing any advanced Green function in the field equations with its retarted counterpart by hand, since both types of Green functions naturally emerge in the variation of the action \cite{Belgacem_2018}.

Among the infinite-derivatives gravity theories, the ones dealing with fractional operators, meaning non-local operators containing a non-integer number of derivatives and in particular fractional powers of the d'Alembertian $(-\Box)^s$, occupy a special position (see e.g. \cite{gorenflo2008fractional} for a review of fractional calculus); since these operators are non analytic, they introduce a stronger departure from local gravity than non-local operators constructed with infinite derivatives of integer order. Once quantized, these kind of theories have been shown to be unitary and finite at one loop for particular values of the fractional exponent $s$ \cite{Calcagni_2021}.

Recently \cite{Benetti_2023,Benetti_2023II,Benetti_2023III} put forward a framework where the dark matter (DM) component within virialized halos is subject to a non-local interaction originated by fractional gravity (FG) effects. We have shown that such a framework can substantially alleviate the small-scale issues of the standard cold dark matter paradigm, without altering the DM mass profile predicted by gravity-only $N-$body simulations (not including possible baryonic feedback effects, which can improve the agreements with observations in large spirals, but can be neglected in very small, dark matter dominated, dwarfs; e.g., \cite{Penarrubia12,Orkney21,Asencio22,Errani24}), while retaining its successes on large cosmological scales \footnote{We caveat the reader that the standard cosmological model faces some challenges on large scales, like the Hubble (e.g., \cite{Riess22,Riess24}), the cosmic shear (e.g., \cite{Asgari21,Secco22}), the bulk flows (e.g., \cite{Watkins23,Whitford23}) and the large local supervoid (e.g., \cite{Haslbauer20,Wong22}) tensions. However, addressing these is beyond the scope of the present work.}. In particular, we were able to: (i) provide accurate fits to the stacked rotation curves of spiral galaxies with different properties, to the thermodynamic profiles of galaxy clusters, and to the rotation curves of individual dwarf spheroidal and irregular galaxies; (ii) reproduce the observed shape and scatter of the radial acceleration relation (RAR) over an extended range of galaxy accelerations; (iii) account for several scaling relations observed between the properties of DM halo and baryonic disk. The emergent evidence is that, when endowed with the fractional interaction, DM performs better that in the Newtonian case, especially in small structures like dwarfs where fractional gravity effects are found to be stronger. Furthermore the presence of DM, whose existence, although not yet confirmed by direct experiments, has been strongly supported by a plethora of astrophysical and cosmological probes \cite{Rubin_1980,Brout_2022,Markevitch_2004,Mendel_2020, Zhao_2022,Planck2018,Bennett_2003}, is crucial since 
we have verified in \cite{Benetti_2023} that a MOND-like version of the fractional model including only baryons is not able to fit the aforementioned data (though the classical, phenomenological MOND parameterization works better; e.g., \cite{Famaey12,Banik22}). Specifically, in such a MOND-like model the outer part of the rotation curves always tend to decrease towards zero, while data show that the general trend is to stabilize around a constant value, or even to increase in the smaller dwarfs. Furthermore, the correct normalization of the rotation curve (even in the inner region) can be obtained only with disk masses systematically larger than those inferred from photometry ($I-$band), hinting to mass-to-light ratios substantially larger than the typical values $M_{\star}/L_{I} \leq 1.5$ expected on the basis of stellar population synthesis models \cite{Portinari2004}.

In fractional gravity the potential originated by a given DM distribution is computed from a modified Poisson equation \cite{GiustiMOND}, in which the Laplace-Beltrami operator is replaced by its fractional power \cite{Benetti_2023,Fractional_Laplacian}
\begin{equation} \label{FPE}
(-\Delta)^s\Phi(\textbf{x}) = -4\pi\, G\, \ell^{2-2s}\, \rho(\textbf{x})\;.
\end{equation}
Here the fractional exponent $s$, which measures the strength of non locality, is restricted to the range $s\in[1,3/2]$ in order to ensure convergence, while the length scale $\ell$, which must be introduced for dimensional consistency, can be interpreted as the typical size below which gravitational effects are somewhat reduced and above which they are instead amplified by non-locality. One can also transfer the non-locality from the differential operator to the matter content and interpret Equation \eqref{FPE} as a classical Poisson equation sourced by an effective, delocalized density distribution $\rho_{\rm eff}(\textbf{x}) = (-\ell^2\,\Delta)^{1-s}\, \rho(\textbf{x})$, or equivalently $\tilde{\rho}_{\rm eff}(\textbf{k}) = (\ell\,|\textbf{k}|)^{2-2s}\, \tilde{\rho}(\textbf{k})$ in the Fourier domain. Considering for instance a point-like particle at the origin $\rho(\textbf{x}) = m\,\delta^{3}(\textbf{x})$, the effective density for this system turns out to be
\begin{equation}
\rho_{\rm eff}(\textbf{x}) = \frac{\Gamma\left(\frac{5}{2}-s\right)}{4^{s-2}\,\sqrt{\pi}\,\Gamma(s-1)}\, \frac{\ell^{2-2s}\, G\, m}{|\textbf{x}|^{5-2s}}\;,
\end{equation}
showing that the effective density associated to a well localized matter distribution can be non-zero in all of space. If both baryons and DM are present with densities $\rho_{\rm bar}$ and $\rho_{\rm DM}$, the potential in Fractional Gravity can then be computed by solving the Poisson equation
\begin{equation} \label{NFG}
    \Delta\Phi(\textbf{x}) =4\pi \, G\, [\rho_{\rm bar}(\textbf{x}) + \rho_{\rm DM, eff}(\textbf{x})]= 4\pi \, G\, [\rho_{\rm bar}(\textbf{x}) + (-\ell^2\, \Delta)^{1-s}\,\rho_{\rm DM}(\textbf{x})] \;.
\end{equation}

Motivated by the successes of Fractional Gravity at the Newtonian level, in this paper we propose a general relativistic extension of the latter, deriving the equations of motion from an action which naturally extends the Einstein-Hilbert one of GR. This will enable us to explore Fractional Gravity in regimes in which the strictly Newtonian approximation cannot be applied, to study phenomena which are exclusive to the relativistic setting (e.g., propagation of gravitational waves), and provide a more straightforward physical interpretation of the non-locality effects associated to the theory.

The plan of the paper is as follows. In Section \ref{sec|action} we introduce the action of the theory and derive the associated equations of motion; in Section \ref{sec|weak} we investigate the weak field limit, derive the Newtonian (Section \ref{sec|Newton}) and post-Newtonian (Section \ref{sec|postNewton}) approximations, and discuss mode propagation via a scalar-vector-tensor decomposition analysis (Section \ref{sec|SVT}); in Section \ref{sec|conservation} we derive the modified conservation equation for the DM fluid; finally, in Section \ref{sec|summary} we summarize our findings and outline future perspectives.

\section{Relativistic action and equations of motion}\label{sec|action}

In extending DM in Fractional Gravity to a relativistic setting, we aim to meet a series of requirements: (i) our theory must be generally covariant and locally Lorentz invariant; (ii) baryons must obey standard GR, in particular must follow the geodesics of the metric obtained by solving the field equations; (iii) the Newtonian limit of the field equations of the relativistic theory must yield the fractional Poisson Equation \eqref{FPE}. Moreover, the expression of the effective DM density suggests to consider negative fractional powers of the d'Alembertian acting on the matter distribution.

Inspired by these guidelines and by the principle of simplicity, we propose an action where the scalar curvature couples with the trace of the DM stress-energy tensor via a non-local operator $\mathcal{F}(\Box)$ which must be chosen in such a way to recover the appropriate Newtonian limit. The action of this model, from now on called Relativistic Scalar Fractional Gravity (RSFG), reads\footnote{We remark that the choice of this action is based on simplicity principles. In fact, other possibilities would include a non-local coupling $R_{\mu\nu}\,\mathcal{F}_2(\Box)\, T_{\rm DM}^{\mu\nu}$ between the Ricci and the DM stress-energy tensors, or even $R_{\mu\alpha\nu\beta}\,\mathcal{F}_4(\Box)\, T_{\rm DM}^{\mu\nu}\, T_{\rm DM}^{\alpha\beta}$ between the Riemann (or Weyl) and two copies of the DM stress-energy tensors. We plan to explore these alternative, more complex theories in a forthcoming paper.}
\begin{align}\label{FGaction}
        S_{\rm RSFG}[g,\psi_{\rm bar},\psi_{\rm DM}] & = S_{\rm EH}[g] + S_{\rm bar}[g,\psi_{\rm bar}] + S_{\rm DM}[g,\psi_{\rm DM}] +\nonumber\\
        & \nonumber \\
        &- \frac{1}{2}\int_{\mathcal{M}}{\rm d}^4x\; \sqrt{-g}\,R\, \mathcal{F}(\Box)\, T_{\rm DM}
    \end{align}    
where $S_{\rm EH}[g] = (2\kappa)^{-1}\,\int_{\mathcal{M}}{\rm d}^4x\;\sqrt{-g}\, R$, with $\kappa\equiv 8\pi\,G/c^4$, is the usual Einstein-Hilbert action of GR, $S_{\rm bar}[g,\psi_{\rm bar}]$ and $S_{\rm DM}[g,\psi_{\rm DM}]$ denote the matter actions for baryons and DM, and $T_{\rm DM}\equiv g^{\mu\nu}\, T_{\rm DM\,\mu\nu}$ is the trace of the DM stress-energy tensor $T_{\rm DM\,\mu\nu}$. We anticipate that the Fractional Poisson Equation considered in our previous works \cite{Benetti_2023,Benetti_2023II,Benetti_2023III} can be recovered in the Newtonian limit if $\mathcal{F}(\Box)$ is chosen of the form
\begin{equation}\label{operator}
\mathcal{F}(\Box) = \ell^2\, [(\ell^2\, \Box)^{-1}+(-\ell^2\, \Box)^{-s}]\;;
\end{equation}
hereafter we will assume this shape and work in units for which $\ell=1$, being the appropriate power of $\ell$ easily reinserted when needed.

To obtain the field equations from the action of Equation \eqref{FGaction}, we vary it with respect to the inverse metric, finding
\begin{align}\label{FGactionvary}
	   \delta S_{\rm RSFG} & = \delta S_{\rm EH}+ \delta S_{\rm bar}+ \delta S_{\rm DM} - \frac{1}{2}\int_{\mathcal{M}} {\rm d}^4x\; \{ \delta(\sqrt{-g})\,R\,\mathcal{F}\, T_{\rm DM} + \sqrt{-g}\, \delta R\, \mathcal{F}\, T_{\rm DM} +\nonumber \\
        &\nonumber \\
        & + \sqrt{-g}\, R\, \left(\delta(\Box^{-1})+\delta[(-\Box)^{-s}]\right)\, T_{\rm DM} + \sqrt{-g}\,R\,\mathcal{F}\,\delta T_{\rm DM}\}=
        \nonumber \\
        &\nonumber \\
        & = \frac{1}{2}\int_{\mathcal{M}} {\rm d}^4 x\; \left\{\sqrt{-g}\,\delta g^{\mu\nu}\,\left[\frac{G_{\mu\nu}}{\kappa} - T_{\rm bar, \mu\nu} -  T_{\rm DM,\mu\nu} - (G_{\mu\nu}+g_{\mu\nu}\,\Box-\nabla_{\mu}\,\nabla_{\nu})\,\mathcal{F}\, T_{\rm DM}+ \nonumber \right. \right. \\
        &\nonumber \\
		& \left. \left. - \mathcal{F}\, R\,\frac{\delta T_{\rm DM}}{\delta g^{\mu\nu}}\right] - \sqrt{-g}\,R\,(\delta(\Box^{-1})+\delta[(-\Box)^{-s}])\,T_{\rm DM}\,\right\}+ \nonumber \\
        &\nonumber \\
        & + \frac{\epsilon}{2\kappa}\oint_{\partial\mathcal{M}}{\rm d}^{3}y\;\sqrt{h}\,N_{\alpha}\, \left[
        g_{\mu\nu}\,(\nabla^{\alpha}\,\delta g^{\mu\nu}-\nabla^{\nu}\,\delta g^{\mu\alpha})\,(1-\kappa\,\mathcal{F}\, T_{\rm DM}) \nonumber \right.+ \nonumber \\
        &\nonumber \\
        & \left. +\kappa\,\delta g^{\alpha\beta}\,\nabla_{\beta}\,\mathcal{F}\, T_{\rm DM} -\kappa\,\delta g^{\mu\nu}g_{\mu\nu}\,\nabla^{\alpha}\,\mathcal{F}\, T_{\rm DM}\right]\;,
	\end{align}
where we have used the standard definition of the stress-energy tensor $T_{\mu\nu} = -\frac{2}{\sqrt{-g}}\frac{\delta S}{\delta g^{\mu\nu}}$, the well-known expression for the variation of the Ricci scalar
\begin{equation}
\delta R = (R_{\mu\nu} + g_{\mu\nu}\,\Box - \nabla_{\mu}\,\nabla_{\nu})\,\delta g^{\mu\nu}\;,
\end{equation}
and we have applied the divergence theorem twice to transfer the action of the differential operators on the stress-energy tensor. The last term in Equation \eqref{FGactionvary} involves an integration over the three dimensional boundary $\partial\mathcal{M}$, whose volume form is ${\rm d}S_{\alpha} = \epsilon\,\sqrt{h}\,N_{\alpha}\,{\rm d}^{3}y$; here $N^{\alpha}$ denotes the normal vector to the surface and $\epsilon \equiv N^{\alpha}N_{\alpha} = \pm 1$ depends on the surface being spacelike or timelike, respectively. 

In line with the standard cosmological model, we consider DM as a perfect, pressureless fluid, with $T^{\mu\nu}_{\rm DM}=\rho\, u^{\mu}\,u^{\nu}$ and $T_{\rm DM} = -\rho$. In this case \cite{haghani2023variation} have shown that, in standard astrophysical and cosmological settings, where the entropy and particle production rates remain unchanged during the dynamical evolution of the fluid, the variation of the trace of the stress energy tensor can be written in a closed form as
    \begin{equation}
        \frac{\delta T_{\rm DM}}{\delta g^{\mu\nu}} = \frac{1}{2}\,(T_{\rm DM}\,g_{\mu\nu}-T_{\rm DM\,\mu\nu})\;.
    \end{equation}
    
To proceed further one needs the variation of the nonlocal operators. This can be found by recalling the semi-group integral representation of the inverse Fractional d'Alembertian \cite{Calcagni_2021}
\begin{equation}
(-\Box)^{-s} = \frac{1}{\Gamma(s)}\,\int^{\infty}_{0}{\rm d}\tau\;\tau^{s-1}\,e^{\tau\Box}
\end{equation}
which stems from the definition of the Euler Gamma function $\Gamma(s)$, together with the Duhamel's formula for the variation of the exponential of an operator $\mathcal{O}$ in terms of the variation of the operator itself
\begin{equation}
\delta(e^{\tau\mathcal{O}})= \int^{\tau}_{0}{\rm d}q\,e^{q\mathcal{O}}\,\delta\mathcal{O}\,e^{(\tau-q)\mathcal{O}}\;.
\end{equation}
In particular, the variation of the inverse d'Alembertian simply follows
\begin{equation}
    \delta\Box^{-1} = -\Box^{-1}\,\delta\Box\,\Box^{-1}\;.
\end{equation}

Therefore, one has
\begin{align*}
    & \int_{\mathcal{M}}{\rm d}^{4}x\; \sqrt{-g}\, R\,\delta(-\Box^{-s})\, T_{\rm DM} = \nonumber \\
        &\nonumber \\
	& = \int_{\mathcal{M}}{\rm d}^{4}x\; \sqrt{-g}\, \frac{1}{\Gamma(s)}\, \int^{\infty}_{0}{\rm d}\tau\,\tau^{s-1}\,\int^{\tau}_{0}{\rm d}q\; R\, e^{q\Box}\, \delta\Box\, e^{(\tau-q)\Box}\,T_{\rm DM} = \nonumber \\
    &\nonumber \\
    & = \frac{1}{\Gamma(s)}\int^{\infty}_{0}{\rm d}\tau\; \tau^{s-1}\, \int^{\tau}_{0}{\rm d}q\; \int_{\mathcal{M}}{\rm d}^4x\;\sqrt{-g}\, e^{q\Box}R\,\left[\delta\left(\frac{1}{\sqrt{-g}}\right)\,\partial_{\alpha}\,(\sqrt{-g}\, g^{\alpha\beta}\, \partial_{\beta})+ \right. \nonumber \\
    &\nonumber \\
    & \left. + \frac{1}{\sqrt{-g}}\,
	\partial_{\alpha}\,(\delta(\sqrt{-g}\, g^{\alpha\beta})\, \partial_{\beta})\right]\, e^{(\tau-q)\Box} T_{\rm DM} = \nonumber \\
  \end{align*}
\begin{align}
        & = \frac{1}{\Gamma(s)}\, \int^{\infty}_{0}{\rm d}\tau\; \tau^{s-1}\, \int^{\tau}_{0}{\rm d}q\; \int_{\mathcal{M}}{\rm d}^4x\; \sqrt{-g}\, e^{q\Box} R\, \left[\frac{1}{2}\, g_{\mu\nu}\, \delta g^{\mu\nu}\, \frac{1}{\sqrt{-g}}\, \partial_{\alpha}\, (\sqrt{-g}\, g^{\alpha\beta}\, \partial_{\beta})+ \right. \nonumber \\ 
        &\nonumber \\
  & \left.+ \frac{1}{\sqrt{-g}}\, \partial_{\alpha}\, \left(\sqrt{-g}\, \delta g^{\mu\nu}\, \left(\delta^{(\alpha}_{\mu}\, \delta^{\beta)}_{\nu}-\frac{1}{2}\, g_{\mu\nu}\, g^{\alpha\beta}\right)\, \partial_{\beta}\right)\right]\, e^{(\tau-q)\Box} T_{\rm DM}  =\nonumber \\
  &\nonumber \\
        & = \frac{1}{\Gamma(s)}\, \int^{\infty}_{0}{\rm d}\tau\; \tau^{s-1}\, \int^{\tau}_{0}{\rm d}q\; \int_{\mathcal{M}}{\rm d}^4x\; \sqrt{-g}\, \delta g^{\mu\nu}\, \left[\frac{1}{2}\, g_{\mu\nu}\, e^{q\Box} R\, e^{(\tau-q)\Box}\, \Box T_{\rm DM} +\nonumber \right. \\
        &\nonumber \\
        & \left. +  \left(\frac{1}{2}\, g_{\mu\nu}\, g^{\alpha\beta}-\delta^{(\alpha}_{\mu}\, \delta^{\beta)}_{\nu}\right)\, \nabla_{\alpha}\, (e^{q\Box} R)\, \nabla_{\beta}(e^{(\tau-q)\Box} T_{\rm DM})\right] +\nonumber \\
        &\nonumber \\
		& + \frac{1}{\Gamma(s)}\int^{\infty}_{0}{\rm d}\tau\; \tau^{s-1}\, \int^{\tau}_{0}{\rm d}q\; \int_{\mathcal{M}}{\rm d}^4x\; \partial_{\alpha}\, \left[ e^{q\Box} R\, \left(\delta^{(\alpha}_{\mu}\, \delta^{\beta)}_{\nu}-\frac{1}{2}\, g_{\mu\nu}\, g^{\alpha\beta}\right)\, \partial_{\beta}\, (e^{(\tau-q)\Box}T_{\rm DM})\, \sqrt{-g}\, \delta g^{\mu\nu}\right]\; .
    \end{align}
Similarly, 
   \begin{align}
	& \int_{\mathcal{M}}{\rm d}^4x\; \sqrt{-g}\, R\, \delta(\Box^{-1})\, T_{\rm DM} = \nonumber \\
 &\nonumber \\
	& = \int_{\mathcal{M}}{\rm d}^4x\; \sqrt{-g}\, \delta g^{\mu\nu}\, \left[-\frac{1}{2}\, g_{\mu\nu}\, T_{\rm DM}\, \Box^{-1}R +  \partial_{\alpha}\,(\Box^{-1}R)\, \left(\delta^{(\alpha}_{\mu}\, \delta^{\beta)}_{\nu}-\frac{1}{2}\, g_{\mu\nu}\, g^{\alpha\beta}\right)\, \partial_{\beta}(\Box^{-1}T_{\rm DM})\right] + \nonumber \\
 &\nonumber \\
	& - \int_{\mathcal{M}}{\rm d}^4x\; \partial_{\alpha}\, \left[ \Box^{-1}R\, \left(\delta^{(\alpha}_{\mu}\, \delta^{\beta)}_{\nu}-\frac{1}{2}\, g_{\mu\nu}\, g^{\alpha\beta}\right)\, \partial_{\beta}\, (\Box^{-1}T_{\rm DM})\, \sqrt{-g}\, \delta g^{\mu\nu}\right]\;.
	\end{align}
The last line of this expression can be turned into a surface term by using the identity $\partial_{\alpha}\,(\sqrt{-g}\, V^{\alpha}) = \sqrt{-g}\, \nabla_{\alpha}\,V^{\alpha}$, valid for any vector $V^{\alpha}$, and the divergence theorem. 

Collecting all the terms together, the field equations read
    \begin{align}\label{RSFGEQ}
		\frac{G_{\mu\nu}}{\kappa} & = T_{\rm bar\,\mu\nu} + T_{\rm DM\,\mu\nu} + (G_{\mu\nu}+g_{\mu\nu}\,\Box-\nabla_{\mu}\,\nabla_{\nu})\,\mathcal{F} \,T_{\rm DM} - \frac{1}{2}\, \mathcal{F}\, R\,(T_{\rm DM\,\mu\nu}-T_{\rm DM}\,g_{\mu\nu})+ \nonumber \\
  &\nonumber \\
  & -\frac{1}{2}\, g_{\mu\nu}\, T_{\rm DM}\, \Box^{-1}R + \left(\delta^{(\alpha}_{\mu}\,\delta^{\beta)}_{\nu}-\frac{1}{2}\, g_{\mu\nu}\, g^{\alpha\beta}\right)\, \nabla_{\alpha}\, (\Box^{-1}R)\, \nabla_{\beta}(\Box^{-1}T_{\rm DM})+ \nonumber \\
  &\nonumber \\
		& -\frac{1}{\Gamma(s)}\, \int_{0}^{\infty}{\rm d}\tau\;\tau^{s-1}\, \int_{0}^{\tau}{\rm d}q\; \left[\left(\delta^{(\alpha}_{\mu}\, \delta^{\beta)}_{\nu}-\frac{1}{2}\, g_{\mu\nu}\, g^{\alpha\beta}\right)\, \nabla_{\alpha}\,\left(e^{q\Box}R\right)\, \nabla_{\beta}\, \left(e^{(\tau-q)\Box}T_{\rm DM}\right)\right.+ \nonumber\\
  &\nonumber \\
  & \left.- \frac{1}{2}\, g_{\mu\nu}\, e^{q\Box}R\, \Box\,  e^{(\tau-q)\Box}T_{\rm DM}\right]\;,
	\end{align}
and the full boundary term takes the form
    \begin{align}
        & \frac{\epsilon}{2\kappa}\,\oint_{\partial\mathcal{M}}{\rm d}^{3}y\; \sqrt{h}\, N_{\alpha}\, \left\{
        g_{\mu\nu}\, (\nabla^{\alpha}\,\delta g^{\mu\nu}-\nabla^{\nu}\,\delta g^{\mu\alpha})\, (1-\kappa\,\mathcal{F}\, T_{\rm DM}) + +\kappa\,\delta g^{\alpha\beta}\,\nabla_{\beta}\,\mathcal{F}\, T_{\rm DM} + \nonumber \right. \nonumber \\
        &\nonumber \\
        & \left. - \kappa\,\delta g^{\mu\nu}g_{\mu\nu}\,\nabla^{\alpha}\,\mathcal{F}\, T_{\rm DM} -  \kappa\, \Box^{-1}R\, \left(\delta^{(\alpha}_{\mu}\, \delta^{\beta)}_{\nu}-\frac{1}{2}\, g_{\mu\nu}\, g^{\alpha\beta}\right)\, \nabla_{\beta}(\Box^{-1}T_{\rm DM})\, \delta g^{\mu\nu} +  \right. \nonumber \\
        &\nonumber \\
        & \left. + \frac{\kappa}{\Gamma(s)}\, \int^{\infty}_{0}{\rm d}\tau\; \tau^{s-1}\, \int^{\tau}_{0}{\rm d}q\; e^{q\Box}R\, \left(\delta^{(\alpha}_{\mu}\, \delta^{\beta)}_{\nu}-\frac{1}{2}\, g_{\mu\nu}\, g^{\alpha\beta}\right)\, \nabla_{\beta}\, (e^{(\tau-q)\Box}T_{\rm DM})\, \delta g^{\mu\nu}\right\}\;.
    \end{align}
    
Although the last three addenda can be set to zero at the boundary since the variation of the metric vanishes there, the situation is more subtle for the first one, which involves the variation of the derivative of the metric at the boundary, and hence cannot be set to zero in principle. However, it is well known that such a term can be related to the variation of the trace of the intrinsic curvature $K \equiv \nabla_{\alpha}\, N^\alpha$ of the boundary. In fact, one can write
    \begin{equation}
        N_{\alpha}\, g_{\mu\nu}\, (\nabla^{\alpha}\, \delta g^{\mu\nu}-\nabla^{\nu}\, \delta g^{\mu\alpha}) = N_{\alpha}\, P_{\mu\nu}\, (\nabla^{\alpha}\, \delta g^{\mu\nu}-\nabla^{\nu}\, \delta g^{\mu\alpha}) = N_{\alpha}\, P_{\mu\nu}\, \nabla^{\alpha}\, \delta g^{\mu\nu}\;,
    \end{equation}
where we have substituted the metric with the projection tensor on the boundary $P_{\mu\nu} \equiv g_{\mu\nu}-\epsilon\,N_{\alpha}\, N_{\beta}$ since the term involving the product of three tangent vectors vanishes by symmetry, and in the last equality we have used the fact that $P_{\mu\nu}\, \nabla^{\nu}\, \delta g^{\mu\alpha}=0$ since it is the covariant derivative of the metric variation projected on the boundary, on which the metric is held fixed. On the other hand,     
\begin{align}
        \delta K & = P^{\alpha}_{\beta}\, \delta \Gamma^{\beta}_{\alpha\gamma}\, N^{\gamma} = -\frac{1}{2}\, N^{\gamma}\, P^{\alpha}_{\beta}\, g^{\beta\delta}\, (\nabla_{\alpha}\, \delta g_{\gamma\delta} + \nabla_{\gamma}\, \delta g_{\alpha\delta} - \nabla_{\delta}\, \delta g_{\gamma\alpha}) = \nonumber \\
        &\nonumber \\
        & = -\frac{1}{2}\, N^{\gamma}\, P^{\alpha}_{\beta}\, g^{\beta\delta}\, \nabla_{\gamma}\, \delta g_{\alpha\delta} = -\frac{1}{2}\, N_{\alpha}\, P_{\mu\nu}\, \nabla^{\alpha}\, \delta g^{\mu\nu}\;
    \end{align}
so that $N_{\alpha}\, P_{\mu\nu}\, \nabla^{\alpha}\, \delta g^{\mu\nu} = -2\, \delta K$ holds. Thus, we obtain
   \begin{align}
    & \oint_{\partial\mathcal{M}}{\rm d}^{3}y\;\sqrt{h}\,N_{\alpha}\,
        g_{\mu\nu}\, (\nabla^{\alpha}\, \delta g^{\mu\nu}-\nabla^{\nu}\,\delta g^{\mu\alpha})\, (1-\kappa\,\mathcal{F}\, T_{\rm DM}) = \nonumber \\
        & \nonumber\\        
        & =-2\, \oint_{\partial\mathcal{M}}{\rm d}^{3}y\;\sqrt{h}\,\delta K\, (1-\kappa\, \mathcal{F}\, T_{\rm DM})\;.
    \end{align}
Note that, although $\mathcal{F}(\Box)$ is a non-local operator, as long as the trace of the DM stress-energy tensor vanishes asymptotically on the boundary, so does $\mathcal{F}(\Box)\, T_{\rm DM}$. This should be contrasted to other modified gravity theories, like for example $F(R)$; for these, in order to have a well-defined variational principle, the quantity $F'(R)$ must be required to vanish on the boundary, implying that it represents a new dynamical degree of freedom \cite{Guarnizo_2010}. In the present case of RSFG, the boundary term can be simply accounted for by introducing in the action a Gibbons Hawking York-like counter-term \cite{Gybbons-Hawking,York} of the form:
    \begin{equation}
        \frac{\epsilon}{\kappa}\, \oint_{\partial\mathcal{M}}{\rm d}^{3}y\; \sqrt{h}\, K\, (1-\kappa\, \mathcal{F}\, T_{\rm DM})\;.
    \end{equation}
Furthemore, one can renormalize the action by subtracting the extrinsic curvature on the boundary of the corresponding asymptotic vacuum spacetime $K_{0}$. 

Therefore the full action of RSFG reads
    \begin{align}
        S_{RSFG}[g,\psi_b,\psi_{\rm DM}] & = S_{EH}[g] + S_{\rm bar}[g,\psi_b] + S_{\rm DM}[g,\psi_{\rm DM}] + \nonumber \\
        &\nonumber \\
        & - \frac{1}{2}\, \int_{\mathcal{M}}{\rm d}^4x\;\sqrt{-g}\,R\,\mathcal{F}\,T_{\rm DM}+\frac{\epsilon}{\kappa}\,\oint_{\partial\mathcal{M}}{\rm d}^{3}y\;\sqrt{h}\, (K-K_{0})\, (1-\kappa\,\mathcal{F}\, T_{\rm DM})\;.
    \end{align}

\section{Weak field limit}\label{sec|weak}

Since the full field Equations \eqref{RSFGEQ} are very complicated, it is convenient to restrict further analysis to the weak field limit, as it is customary in non-local gravity theories (see, e.g., \cite{buoninfante_2022black}). To this purpose, one assumes the dynamics of the gravitational field to be encoded in a small perturbation over the Minkowski background $g_{\mu\nu} = \eta_{\mu\nu} + h_{\mu\nu}$,  with $h_{\mu\nu}\ll 1$. 
Furthermore, the non-local operator $\mathcal{F}(\Box)$ is obtained by replacing the D'Alembertian with its expression in Minkowksi space $\mathcal{F}(\Box_{g})\simeq \mathcal{F}(\Box_{\eta})$, to linear order in $h$. Finally, having the metric as a small perturbation requires the stress energy tensor to be at least linear in $h$, so that all the terms in Equation \eqref{RSFGEQ} involving the product of the curvature and the stress-energy tensor can be neglected at the linear order. With these assumptions Equation \eqref{RSFGEQ}, to linear order in $h$, simplifies considerably and read
    \begin{align} \label{WFL}
        \frac{{G}^{(h)}_{\mu\nu}}{\kappa}= T_{\rm bar\,\mu\nu} + T_{\rm DM\,\mu\nu} - (\partial_{\mu}\,\partial_{\nu}-\eta_{\mu\nu}\,\Box_{\eta})\, \mathcal{F}(\Box_{\eta})\, T_{\rm DM}\;.
    \end{align}
We note that, in this weak field limit, the contracted Bianchi identities imply that the baryons and DM stress energy tensors must satisfy the usual conservation law $\partial^{\mu}\,(T_{\rm bar\,\mu\nu} + T_{\rm DM\,\mu\nu})=0$, in terms of flat partial derivatives. We will see in the next Section that this is no longer true in RSFG when the higher order terms are included, due to the appearance of an external non-local force.

If we further assume the metric to be quasi static and the source to move slowly with respect to the speed of light $c$, we can expand Equation \eqref{WFL} in inverse powers of $1/c$ to obtain the Newtonian and post-Newtonian approximations. It is customary to express the metric perturbation in terms of potentials, constructed from the matter density and velocity, with real coefficients known as post-Newtonian parameters. In a gravity theory, these must be computed from the field equations, and then allow a direct comparison with experimental data (see \cite{Gravitation}, Chapters 39 and 40, for an excellent discussion on the Parametrized Post-Newtonian, or PPN, expansion and its relevance in performing experimental tests of GR in the Solar System). If we retain only terms up to order $1/c^{2}$ the metric perturbation, in the standard PPN coordinate system, can be written in the form
\begin{equation}\label{standard}
    h_{\mu\nu} = 
    \begin{pmatrix} 
        -\frac{2\Phi}{c^2} &  \\
        & -\frac{2\Psi}{c^2}\,\delta_{ij}  
    \end{pmatrix}\;;
\end{equation}
here $\Phi$ must coincide with the gravitational potential in the Newtonian limit of the theory, while $\Psi$ is usually written as $\Psi = \gamma\,\Phi$ in terms of the PPN parameter $\gamma$, representing the amount of space curvature per unit rest mass. 

\subsection{Newtonian limit: recovering the Fractional Poisson Equation}\label{sec|Newton}

Specializing to RSFG, we can insert Equation \eqref{standard} into the expression of the linearized Einstein tensor 
\begin{equation}
    G_{\mu\nu}^{(h)} = \frac{1}{2}\,[2\,\partial_{\alpha}\,\partial_{(\mu}\,h^\alpha_{\nu)}-\partial_{\mu}\,\partial_{\nu}\,h-\Box h_{\mu\nu}-\eta_{\mu\nu}\,\partial_{\alpha}\,\partial_{\beta}\,h^{\alpha\beta} + \eta_{\mu\nu}\,\Box h]
\end{equation}
to find, up to order $1/c^2$, the expansion:
\begin{equation}
    G_{00}^{(h)} = \frac{2\,\Delta\Psi}{c^2}, \,\,\,\,\, G_{0i}^{(h)} = 0, \,\,\,\,\, G_{ij}^{(h)} = (\partial_{i}\,\partial_{j}-\delta_{ij}\,\Delta)\,\frac{\Psi-\Phi}{c^2}\;.
\end{equation} 
On the other hand, expanding the right hand side of Equation \eqref{WFL} to the same order, we obtain the equations 
\begin{align}
     & \Delta\Psi = 4\pi G\,\rho_{\rm bar} + 4\pi G\,(1+\Delta\,\mathcal{F})\,\rho_{\rm DM} \label{Psi} \\
     & \nonumber\\
     & (\partial_{i}\,\partial_{j}-\delta_{ij}\,\Delta)\,(\Psi-\Phi) = 8\pi G\,(\partial_{i}\,\partial_{j}-\delta_{ij}\,\Delta)\,\mathcal{F}\,\rho_{\rm DM}\;, \label{eq|difference}
\end{align}
with $\rho$ denoting the rest mass density.
To find the connection with the Newtonian potential appearing in the Fractional Poisson Equation \eqref{FPE} of our previous works \cite{Benetti_2023,Benetti_2023II,Benetti_2023III}, we recall that the latter must be encoded in the combination $\Delta\Phi_s/c^2 = R_{00} = (G_{00}+\delta^{ij}G_{ij})/2$, to guarantee that the Newtonian limit of the geodesic equation is consistent with the Newtonian equation of motion for the matter. This yields the expected result $\Phi = \Phi_s$, and 
    \begin{equation}
        \Delta\Phi_s = 4\pi G\,\rho_{\rm bar} + 4\pi G\, (1-\Delta\, \mathcal{F})\,\rho_{\rm DM}\;.
    \end{equation}
By taking the Newtonian limit of the non-local operator $\mathcal{F}$ after Equation \eqref{operator}, one has
\begin{equation}
    \mathcal{F} = \Delta^{-1} + \ell^{2-2s}\,(-\Delta)^{-s}
\end{equation}
and hence the Fractional Poisson Equation \eqref{NFG} 
\begin{equation}
    \Delta\Phi_s(\textbf{x}) = 4\pi G\,\rho_{\rm bar}(\textbf{x}) + 4\pi G\, \ell^{2-2s}\, (-\Delta)^{1-s}\, \rho_{\rm DM}(\textbf{x})\;
\end{equation}
is recovered in this limit.

\subsection{Post-Newtonian limit: the parameter $\gamma$}\label{sec|postNewton}

The other potential is straightforwardly found from Equation \eqref{Psi} as
\begin{equation}
    \Psi = 2\,\Phi_{\rm N} - \Phi_s\;,
\end{equation}
where $\Phi_{\rm N}$ denotes the potential in the Newtonian theory generated by the total density distribution $\rho_{\rm bar} + \rho_{\rm DM}$, hence the potential satisfying the standard Poisson equation. The expressions of the post-Newtonian potentials have two remarkable consequences. First, the post-Newtonian parameter $\gamma$ is not constant with radius, and explicitly given by 
\begin{equation}
    \gamma = 2\,\frac{\Phi_{\rm N}}{\Phi_s}-1\;,
\end{equation}
at variance with GR, where instead $\Phi = \Psi = \Phi_{\rm N}$ and hence $\gamma = 1$ applies. Second, weak gravitational lensing in RSFG is not altered with respect to GR, since it depends only on the combination $(\Phi + \Psi)/2 = \Phi_N$ (see e.g. \cite{Gravitation}, Chapter 40; also \cite{zamani2024gravitational}). 

We stress that these two findings are not in contradiction since the Newtonian limit of RSFG boils down to the fractional (not the Newtonian) Poisson equation; other modified gravity theories which have the standard Poisson equation as Newtonian limit imply that lensing is unaffected only if $\gamma=1$; e.g., this is the case in Chameleon gravity \cite{Chameleon}. In parallel, this occurrence allows RSFG to escape the weak lensing constraints $\gamma-1 \approx (2.1 \pm 2.3)\times 10^{-5}$ from the \textit{Cassini} experiment \cite{Cassini}.

The occurrence $\Phi\neq\Psi$ in RSFG is easily understood by noticing that the right hand side of Equation \eqref{WFL} can be interpreted as an effective stress energy tensor for a fluid with both isotropic and anisotropic stresses:
    \begin{equation}\label{ASET}
        \bar{T}_{\rm DM\, \mu\nu} = (\bar{\rho}+\bar{p})\,u_{\mu}\,u_{\nu} + \bar{p}\,\eta_{\mu\nu} + \bar{\Pi}_{\mu\nu}\;,
    \end{equation}
where $\bar{\Pi}_{\mu\nu}$ encodes deviations from a perfect fluid. The various components can be found by comparison of Equations \eqref{WFL} and \eqref{ASET}, requiring that $\bar{T}_{\rm DM} = -\bar{\rho}+3\bar{p}$, and that $\bar{T}_{\rm DM,00} = \bar{\rho}$, $\bar{\Pi}_{00}=0$ in a comoving inertial frame. The result reads
    \begin{align}
        & \bar{\rho} = (1+\Delta\,\mathcal{F})\,\rho_{\rm DM}, \;\;\;\;  \bar{p} = \left(\partial^2_0 -\frac{2}{3}\,\Delta\right)\,\mathcal{F}\,\rho_{\rm DM}\\
        &\nonumber \\
        & \bar{\Pi}_{\mu\nu} = \left[\left(\partial_{\mu}\,\partial_{\nu}-\frac{1}{3}\,\eta_{\mu\nu}\,\Delta\right)-u_{\mu}\,u_{\nu}\left(\partial^2_0 +\frac{1}{3}\,\Delta\right)\right]\,\mathcal{F}\,\rho_{\rm DM}\;. \label{aniso}
    \end{align}
As it is well known, it is found that the difference between the two scalar potentials is proportional to the anisotropic stress of the source in the comoving frame
    \begin{align}
        \Delta(\Psi-\Phi) & = 12\pi\, G\,\left(\frac{\partial_{i}\,\partial_{j}}{\Delta}-\frac{1}{3}\,\delta_{ij}\right)\,\bar{\Pi}^{ij} = \nonumber\\
                &\nonumber\\
        & = 12\pi\, G\,\left(\frac{\partial_{i}\,\partial_{j}}{\Delta}-\frac{1}{3}\,\delta_{ij}\right)\,\left(\partial^{i}\,\partial^{j}-\frac{1}{3}\,\delta^{ij}\,\Delta\right)\,\mathcal{F}\,\rho_{0,\rm DM} = \nonumber \\
                &\nonumber\\
        & = 8\pi G\,\Delta\,\mathcal{F}\rho_{0,\rm DM} = 2\,\Delta\,(\Phi_{\rm N}-\Phi_s)\;.
    \end{align}
Note that the difference depends only on the DM density, as it should, since baryons generate gravity only through their stress-energy tensor, whose anisotropic component vanishes in the Newtonian limit. 
Remarkably, the effective density appearing in the Fractional Poisson Equation can be rewritten as 
\begin{equation}
\rho_{\rm eff} = (1-\Delta\, \mathcal{F})\,\rho_{\rm DM} = \bar{\rho}+3\,\bar{p}\;,
\end{equation}
which is the same expression appearing for instance in the cosmological FRW metric. We conclude that the deviation from the standard cold DM behavior induced by the non-local interaction can be seen as the action of a non-local pressure, acting as a source of gravity even in the non relativistic regime.

\subsection{SVT decomposition: gravitational waves and propagating scalar modes}\label{sec|SVT}

The weak field limit is also the relevant setting for studying the generation and propagation of gravitational waves and other modes. By looking at Equation \eqref{WFL} it is clear that the only term of RSFG which is not present in GR has a very special form, namely it just involves derivatives of a scalar. 

If we perform a Scalar-Vector-Tensor (SVT) decomposition of Equation \eqref{WFL} by separating the metric and the effective stress energy tensor into irreducible representations of spatial rotations, the gravitational waves (corresponding to pure tensorial components) will not be affected by the scalar fractional interaction, since the scalar, vector and tensor modes do not mix with each others in the linear regime.  In particular, their generation will be due as usual to the pure tensorial part of the stress-energy tensors of both baryons and DM, and their propagation, occurring at the speed of light, will be encoded in the two degrees of freedom of a transverse traceless symmetric tensor, which oscillates orthogonally to the direction of motion.\par
On the other hand, it is clear that the scalar part of the decomposition gets modified when the fractional interaction is taken into account. Indeed, focusing only on the latter, we can write the metric and the total stress energy tensor as
    \begin{align}
        & h_{00} = 2\,\phi,  & T_{00} & = \rho_{\rm bar} + (1+\Delta\,\mathcal{F})\,\rho_{\rm DM} \nonumber \\
        &\nonumber\\
        & h_{0i} = \partial_{i}k, & T_{0i} & = \partial_{i}S + \partial_i\,\partial_0\,\mathcal{F}\,\rho_{\rm DM}  \nonumber \\
        &\nonumber\\
        & h_{ij} = \frac{1}{3}\,H\,\delta_{ij} + \left(\partial_{i}\,\partial_{j}-\frac{1}{3}\,\delta_{ij}\,\Delta\right)\lambda, & T_{ij} & = \left\{\frac{1}{3}\,p+\left(\partial_0^2-\frac{2}{3}\,\Delta\right)\,\mathcal{F}\,\rho_{\rm DM}\right\}\,\delta_{ij}+ \nonumber \\
        & & & + \left(\partial_{i}\partial_{j}-\frac{1}{3}\delta_{ij}\Delta\right)(\sigma+\,\mathcal{F}\,\rho_{\rm DM}) 
    \end{align}
for some scalar functions $\phi, k, H, \lambda, S, \sigma$. By performing the same SVT decomposition on the generators of gauge transformations, one can check that the following Bardeen's variables are gauge invariant \cite{Bardeen}:
    \begin{align}
        & \psi = -\phi + \partial_{0}\,k -\frac{1}{2} \,\partial^{2}_{0}\,\lambda \nonumber \\
                &\nonumber\\
        & \theta = \frac{1}{3}\,(H-\Delta\,\lambda)
    \end{align}
In terms of these the Einstein tensor is given by \cite{Barausse}:
    \begin{align}
        & G_{00}^{(h)} = -\Delta\,\theta \nonumber \\
        &\nonumber\\
        & G_{0i}^{(h)} =  -\partial_{0}\,\partial_{i}\,\theta \nonumber \\
        &\nonumber\\
        & G_{ij}^{(h)} = -\frac{1}{2}\,\partial_{i}\,\partial_{j}\,(2\,\psi+\theta) + \delta_{ij}\,\left[\frac{1}{2}\,\Delta\,(2\,\psi+\theta)-\partial_{0}^{2}\,\theta\right]\;.
        \end{align}
We then substitute this decomposition into Equation \eqref{WFL} and use the conservation law $\partial_{\mu} \,T^{\mu}_{\nu}=0$ to find that some of the equations are automatically satisfied on shell, while the others can be simplified to
    \begin{align}\label{prop}
        & \Delta\,\theta= -8\pi G\,[\rho_{\rm bar}+(1+\Delta\,\mathcal{F})\,\rho_{\rm DM}] \nonumber \\
        &\nonumber\\
        & \Delta\,\psi = 4\pi G\,[\rho_{\rm bar} + (1-\Delta\,\mathcal{F})\,\rho_{\rm DM} +3\,p-3\,\partial_{0}\,S]\;.
    \end{align}
When the non local interaction is neglected, the conservation of the energy momentum tensor implies that these degrees of freedom do not propagate, leaving as the only propagating modes the pure tensorial ones, representing gravitational waves. However, in the current framework, this is not the case, since the presence of $\mathcal{F}(\Box)$, containing inverse powers of the d'Alembertian, implies that these scalar modes do propagate. In fact, the solution of \eqref{prop} for $\theta$ is
\begin{align}\label{theta}
    \theta(x,t) & = 2\,G\,\int{\rm d}^3x'\;\frac{\rho_{\rm bar}(x',t)+\rho_{\rm DM}(x',t)}{\lvert x-x'\rvert} +  2\,G\,\int{\rm d}^3x'\;\frac{\rho_{\rm DM}(x',t-\lvert x-x'\rvert)}{\lvert x-x'\rvert}+ \nonumber \\
    &\nonumber\\
    & -G\,\int{\rm d}^4x'\;\rho_{\rm DM}(x',t')\,\frac{[(t-t')^2-|x-x'|^2]^{s-2}}{4^{s-2}\,\Gamma(s)\,\Gamma(s-1)}\,\Theta(t-t'-\lvert x - x'\rvert)\;,
\end{align}
and an analogue solution holds for $\psi$. The above expression comprises three terms: the first one is the instantaneous Newtonian solution, the second one represents the generation of a scalar wave propagating at the speed of light in vacuum, and the last one is of non-local nature. The Heaviside function in this last term restricts the integration in the interior of the past light cone, thus preventing acausal or superluminal motion (see \ref{app|green} for a derivation of the retarded Green function for the fractional d'Alembertian with $s>1$ in Minkowskian spacetime, whose convolution with the density yields the last term). Note that, if the source is static, the two Bardeen variables are related to the two scalar potentials of the post-Newtonian expansion via $\psi = \Phi_s$, $\theta = -2\Psi$.

The appearance of scalar propagating degrees of freedom alongside gravitational waves is a common feature of many scalar-tensor gravity models which predict a massive or a massless scalar field satisfying a modified Klein-Gordon equation \cite{GW_MassiveGravity,GWF(R),Non-LocalGW,Katsuragawa_2019}. In the present model these scalar modes are associated only with the DM component and feature both a static and a propagating term.

\section{Conservation equations}\label{sec|conservation}

We have shown above that in the weak field limit the stress energy tensor in RSFG satisfies the same local conservation equation valid in GR. However, this is no longer true when  higher order terms are included. In general, the non-local coupling between the DM fluid and gravity modifies the conservation equation by introducing a non-local force, orthogonal to the direction of motion, and depending on the value of the density at any point in spacetime.

To show this occurrence, we start by looking at Equation \eqref{RSFGEQ}, neglecting baryons for simplicity and dropping the subscript 'DM' to simplify the notation. By the contracted Bianchi identities the left hand side satisfy $\nabla_{\mu}\,G^{\mu\nu}=0$, and thus this must also be true for the right hand side. After some algebraic manipulation, in particular by using the relation $[\Box,\nabla_{\mu}]\,\phi = R_{\mu}^{\,\nu}\,\nabla_{\nu}\phi$ valid for any scalar $\phi$, and by integrating by parts with respect to the auxiliary variable $q$ to cancel some terms, we obtain the conservation equation in RSFG 
	\begin{align} \label{Bianchi}
		& 2\,\nabla_{\mu}T^{\mu}_{\nu} + T\,\nabla_{\nu}\,\mathcal{F}\, R -\nabla_{\mu}\,(\mathcal{F}\, R\,T^{\mu}_{\nu}) - \nabla_{\nu}\int_{0}^{\infty}{\rm d}\tau\;\frac{\tau^{s-1}}{\Gamma(s)}\,\int_{0}^{\tau}{\rm d}q\;
		\left[e^{q\Box}\,R\,\Box\, e^{(\tau-q)\Box}\,T\right] = 0\;.
	\end{align}
Including baryons would just amount to add a term $\nabla_\mu T^{\mu}_{\rm bar\,\nu}$ to the left hand side of the above expression.
 
To gain some physical insights, it is useful to decompose the conservation equation in two independent components: one parallel and one orthogonal to the flow of the fluid. The latter can be found by contracting with the tensor $P^{\mu}_{\nu} = \delta^{\mu}_{\nu}+u^\mu\, u_\nu$, which is easily seen to be orthogonal to the fluid velocity $u_\mu\, P^{\mu}_{\nu} = P^{\mu}_{\nu}\,u^\nu=0$. Thus, given any scalar $S$, we can write
    \begin{equation}
        \nabla_\nu S = \delta^{\mu}_{\nu}\,\nabla_\mu S = (P^{\mu}_{\nu}-u^\mu\, u_\nu)\,\nabla_\mu S = P^{\mu}_{\nu}\,\nabla_\mu S - \frac{{\rm d}S}{{\rm d}\mathcal{T}}\,u_\nu\;,
    \end{equation}
where the derivative in the last term is taken with respect the proper time $\mathcal{T}$ along the worldline of the DM fluid element. Furthermore, we have
    \begin{equation}
        \nabla_{\mu}T^{\mu}_{\nu} = u^\mu\,\nabla_{\mu}\,\rho\, u_{\nu} + \rho\,u_\nu\,\nabla_{\mu}\, u^\mu + \rho\, u^\mu\,\nabla_{\mu} \,u_\nu = \left(\frac{{\rm d}\rho}{{\rm d}\mathcal{T}} + \rho\,\theta\right)\,u_\nu + \rho\, a_\nu\;,
    \end{equation}
where $\theta =u^\mu\,\nabla_{\mu}$ is the expansion rate along the direction of the flow of the fluid and $a_\nu$ is the four acceleration, which is orthogonal to the flow. Projecting Equation \eqref{Bianchi} in the two orthogonal directions one obtains two coupled equations:
    \begin{align}\label{Conservation}
        & \left(1-\frac{1}{2}\,\mathcal{F}\, R\right)\,\left(\frac{{\rm d}\rho}{{\rm d}\mathcal{T}} + \rho\,\theta\right) - \frac{1}{2\,\Gamma(s)}\frac{{\rm d}}{{\rm d}\mathcal{T}}\,\int_{0}^{\infty}{\rm d}\tau\;\tau^{s-1}\,\int_{0}^{\tau}{\rm d}q\; \left[e^{q\Box}R\,\Box\, e^{(\tau-q)\Box}\,\rho\right] = 0\;,
        \end{align}
and 
\begin{align} \label{Euler}
        & \left(2-\mathcal{F}\, R\right)\,\rho\, a_\nu - \rho\, P^{\mu}_{\nu}\,\nabla_{\mu}\,\mathcal{F}\, R + P^{\mu}_{\nu}\,\nabla_{\mu}\,\int_{0}^{\infty}{\rm d}\tau\; \frac{\tau^{s-1}}{\Gamma(s)}\,\int_{0}^{\tau}{\rm d}q\;
		\left[e^{q\Box}R\,\Box\, e^{(\tau-q)\Box}\,\rho\right] = 0\;.
    \end{align}
Equation \eqref{Conservation} is the conservation equation describing the evolution of the DM density along its trajectory, while Equation \eqref{Euler} can be interpreted as a modified Euler equation for the DM fluid, where a non-local force appears
    \begin{equation}
        F_\nu = \frac{1}{2-\mathcal{F}\, R}\,\left\{P^{\mu}_{\nu}\,\nabla_{\mu}\mathcal{F}\, R-\frac{1}{\rho}\,P^{\mu}_{\nu}\,\nabla_{\mu}\,\int_{0}^{\infty}{\rm d}\tau\;\frac{\tau^{s-1}}{\Gamma(s)}\,\int_{0}^{\eta}{\rm d}q\;
		\left[e^{q\Box}R\,\Box\, e^{(\tau-q)\Box}\,\rho\right]\right\}\;.
    \end{equation}
The consequence of this force orthogonal to the four velocity is that the DM fluid does not move on the geodesic solutions of the field equations, in accordance with the appearance of a non local pressure in its effective energy momentum tensor. At face value, such a non-local force seems to imply that the gravitational field experienced by a test particle is eventually determined by the overall mass distribution in the Universe, a concept that may conjure to a Mach principle argument \cite{Sciama1953,das2023}; addressing the issue is clearly beyond the scope of the present work, but we plan to come back to it in the next future.

\section{Summary and outlook}\label{sec|summary}

In a series of recent papers \cite{Benetti_2023,Benetti_2023II,Benetti_2023III} we put forward a “fractional gravity" framework which strikes an intermediate course between a modified gravity theory and an exotic dark matter (DM) scenario. It envisages that the DM component in virialized halos is subject to a non-local interaction mediated by gravity. In such a framework the gravitational potential associated to a given DM density distribution is determined by a modified Poisson equation including fractional derivatives, that are aimed at describing non-locality. Remarkably, the dynamics in such a Newtonian fractional gravity setup can be reformulated in terms of the standard Poisson equation, but with an effective density distribution which, especially in small systems like dwarf galaxies, is flatter in the inner region with respect to the true one. We have shown that this occurrence substantially alleviates the small-scale issues of the standard cosmological paradigm, while preserving the DM mass profile predicted by gravity-only $N-$body simulations, and retaining its successes on large cosmological scales. 

Motivated by the above encouraging results, in this work we look for a general relativistic extension of such a framework. Specifically, we propose a theory, dubbed Relativistic Scalar Fractional Gravity or RSFG, in which the trace of the DM stress-energy tensor couples to the scalar curvature via a non-local operator $\mathcal{F}(\Box)$, taken as a function of d'Alembertian to ensure general covariance. 

Our main results can be summarized as follows:

\begin{itemize}

\item We have derived the field equations for RSFG starting from an action principle;

\item We have investigated the weak field limit of RSFG, 
showing that the latter can be represented as GR sourced by an effective DM stress energy tensor, featuring an anisotropic stress of non local nature;

\item We have demonstrated that in the Newtonian limit RSFG reduces to the fractional gravity setup of our previous works if the shape of the non-local operator $\mathcal{F}(\Box) = \Box^{-1} + \ell^2(-\ell^2\Box)^{-s}$ is adopted, with $\ell$ a scale-length and $s$ a fractional index;

\item  We have shown that in the Newtonian limit the deviation of RSFG with respect to the standard Newtonian setup can be interpreted in terms of a non local pressure, which gravitates even in the non relativistic regime;

\item We have analyzed the post-Newtonian approximation and derived the first-order correction parameter $\gamma$. Although the latter deviates from the GR value of unity due to the presence of the aforementioned anisotropic stress in the weak-field limit, weak lensing is not modified with respect to what one finds in GR.

\item We have performed a standard scalar-vector-tensor-decomposition of RSFG in the weak field limit, to highlight that gravitational waves propagate at the speed of light, though also an additional scalar mode becomes dynamical;

\item We have derived the modified conservation laws of the DM stress energy tensor in RSFG, showing that a new non-local force emerges. It implies that the DM fluid  deviate from the geodesic solutions of the field equations. 

\end{itemize}

In future works we plan to: analyze the behavior of RSFG in a standard cosmological setting; look for specific solutions of RSFG in the strong gravity regime; test the predictions of RSFG concerning the post-Newtonian parameter $\gamma$ by designing specific tests in DM-dominated systems;
investigate other terms in the relativistic action that could lead to the same fractional gravity setup in the Newtonian limit, like for example a non-local coupling of the DM stress-energy and the Ricci tensors. 

\section*{Acknowledgments}
We thank the three anonymous referees for very helpful comments and suggestions. We warmly thank M. Adil Butt, C. Baccigalupi, Y. Boumechta, B.S. Haridasu, and S. Silveravalle for useful discussions. This work is partially funded from the projects: ``Data Science methods for MultiMessenger Astrophysics \& Multi-Survey Cosmology'' funded by the Italian Ministry of University and Research, Programmazione triennale 2021/2023 (DM n.2503 dd. 9 December 2019), Programma Congiunto Scuole; EU H2020-MSCA-ITN-2019 n. 860744 \textit{BiD4BESt: Big Data applications for Black hole Evolution Studies}; Italian Research Center on High Performance Computing Big Data and Quantum Computing (ICSC), project funded by European Union - NextGenerationEU - and National Recovery and Resilience Plan (NRRP) - Mission 4 Component 2 within the activities of Spoke 3 (Astrophysics and Cosmos Observations); PRIN MUR 2022 project n. 20224JR28W "Charting unexplored avenues in Dark Matter"; INAF Large Grant 2022 funding scheme with the project "MeerKAT and LOFAR Team up: a Unique Radio Window on Galaxy/AGN co-Evolution; INAF GO-GTO Normal 2023 funding scheme with the project "Serendipitous H-ATLAS-fields Observations of Radio Extragalactic Sources (SHORES)".

\begin{appendix}

\section{Retarded Green function for the fractional d'Alembertian}\label{app|green}

In this appendix we derive the retarded Green function for the fractional d'Alembertian $(-\Box)^s$ with $s>1$ in the case of Minkowksian spacetime, which is exploited in Equation \eqref{theta} of the main text. By definition the Green function satisfies the functional equation
\begin{equation}
	(-\Box_x)^s\,G_{\rm ret}(x-y) = \delta^4(x-y),
\end{equation}
for any two spacetime points $x = (x^0,\textbf{x})$ and $y=(y^0,\textbf{y})$. This equation is best solved recalling the action of the Fractional d'Alembertian in Fourier space; specifically, the expression $\widetilde{(-\Box)^s f}(k) = (k_{\mu}\,k^{\mu})^{s}\,\tilde f(k)$ applies for any function $f(x)$, so that
\begin{equation}
	\tilde{G}_{\rm ret}(k) = (k_{\mu}\,k^{\mu})^{-s}\;.
\end{equation}
Going back to position space, we obtain:
\begin{align}
	G_{\rm ret}(x-y) & = \int\frac{{\rm d}^4\,k}{(2\pi)^4}\;\frac{e^{ik_{\mu}\,(x^{\mu}-y^{\mu})}}{(k_{\nu}\,k^{\nu})^s}= \nonumber \\
    & \nonumber \\
    & = \frac{1}{4\pi^3\,\lvert\textbf{x}-\textbf{y}\rvert}\,\int_{0}^{\infty}{\rm d}k\;k\,\sin(k\,\lvert\textbf{x}-\textbf{y}\rvert)\,\int_{-\infty}^{\infty}{\rm d}k^0\;\frac{e^{-ik^{0}\,(x^0-y^0)}}{[k^2-(k^0+i\epsilon)^2]^s}\;,
\end{align}
with $k=|{\bf k}|$. We then perform a Wick rotation by setting $q = \epsilon - ik^0$, to get:	
\begin{align}\label{appwick}
	G_{\rm ret}(x-y) & = \lim_{\,\epsilon\to 0^+}\,\Theta(x^0-y^0)\,\frac{e^{-\epsilon\,(x^0-y^0)}}{4\pi^3\,i\,\lvert\textbf{x}-\textbf{y}\rvert}\,\int_{0}^{\infty}{\rm d}k\;k\,\sin(k\,\lvert\textbf{x}-\textbf{y}\rvert)\,\int_{\epsilon-i\infty}^{\epsilon +i\infty}{\rm d}q\;\frac{e^{q\,(x^0-y^0)}}{(|\textbf{k}|^2+q^2)^s}= \nonumber \\
&\nonumber\\
 & = \frac{\Theta(x^0-y^0)\,(x^0-y^0)^{s-1/2}}{2^{s+1/2}\,\pi^{3/2}\,\Gamma(s)\,\lvert\textbf{x}-\textbf{y}\rvert}\,\int_{0}^{\infty}{\rm d}k\;k^{3/2-s}\,\sin(k\,\lvert\textbf{x}-\textbf{y}\rvert)\,J_{s-1/2}(k\,(x^0-y^0))\;
\end{align}
where in the last integral of the upper equation we have recognized the inverse Laplace transform of the quantity $(k^2+q^2\,)^{-s}$, and then we have computed it in terms of the Bessel function $J_{\nu}(z)$ of the first kind and order $\nu$. 
To proceed further we can exploit the result that the special integral $I(a,b,\lambda,\nu)\equiv\int_{0}^{\infty}{\rm d}z\;z^{\lambda}\,\sin(bz)\,J_{\nu}(az)$ for $-{\rm Re}(\nu)-1 < {\rm Re}(\lambda) + 1 < 3/2$ can be written in terms of the Gaussian hypergeometric function $_2F_1$ as (e.g., see \cite{gradshteyn2007}, p.730, eq. 6.699)	
\begin{align}
    I(a,b,\lambda,\nu) &=
    \frac{2^{\lambda+1}}{a^{\lambda+2}}\,\frac{\Gamma\left(\frac{2+\lambda +\nu}{2}\right)}{\Gamma\left(\frac{\nu-\lambda}{2}\right)}\,_2F_1\left(\,\frac{2+\lambda +\nu}{2},\,\frac{2+\lambda-\nu}{2};\,\frac{3}{2};\,\frac{b^2}{a^2}\,\right)\;\;,\;\;\;\;{\rm if}\;\;0<b<a \nonumber \\
&\nonumber\\
     & = \left(\frac{a}{2}\right)^{\nu}\,\frac{1}{b^{\lambda+\nu+1}}\,\frac{\Gamma(\lambda+\nu+1)}{\Gamma(\nu+1)}\,\sin\left[\pi\left(\frac{\lambda+\nu+1}{2}\right)\right]\times \nonumber\\
     &\nonumber\\
     & \;\;\;\;\;\;\times \;_2F_1\left(\,\frac{\lambda +\nu+2}{2},\,\frac{\lambda+\nu+1}{2};\,\nu+1;\,\frac{a^2}{b^2}\,\right)\;\;,\;\;\;\;{\rm if}\;\;0<a<b\;.
 \end{align}

In the present context $\lambda = 3/2-s$, $\nu = s-1/2$, $a=x^0-y^0$, $b = \lvert\textbf{x}-\textbf{y}\rvert$ applies, so that $(\lambda+\nu+1)/2 = 1$ holds. This implies that the integral in Equation \eqref{appwick} vanishes whenever $\lvert\textbf{x}-\textbf{y}\rvert > x^0-y^0$, so preventing superluminal motions. Therefore the support of the Green function coincides with the interior of the past light cone for the point $x$, where it is explicitly given by:
\begin{align}
	G_{\rm ret}(x-y) & = \frac{(x^0-y^0)^{2s-4}}{2^{2s-1}\,\pi\,\Gamma(s)\,\Gamma(s-1)}\,_2F_1\,\left(\,\frac{3}{2},\,2-s;\,\frac{3}{2};\,\left[\frac{\lvert\textbf{x}-\textbf{y}\rvert}{x^0-y^0}\right]^2\,\right)\,\Theta(\,x^0-y^0-\lvert\textbf{x}-\textbf{y}\rvert\,) \;.
\end{align}
Finally, since $_2F_1\,\left(\,\frac{3}{2},\,2-s;\,\frac{3}{2};z\right)\, = \left(1-z^2\right)^{s-2}$, we obtain:	
\begin{align}
	G_{\rm ret}(x-y) = \frac{[(x^{0}-y^{0})^2-\lvert\textbf{x}-\textbf{y}\rvert^2]^{s-2}}{2^{2s-1}\,\pi\,\Gamma(s)\,\Gamma(s-1)}\,\Theta(\,x^0-y^0-\lvert\textbf{x}-\textbf{y}\rvert\,)\;,
\end{align}
which is employed in deriving the last term of Equation \eqref{theta} in the main text.
 
\end{appendix}

\section*{References}
\bibliography{mybib}
\bibliographystyle{unsrt}

\end{document}